\pdfoutput=1
\documentclass[journal]{IEEEtran}

\usepackage{amsfonts}
\usepackage{amsmath,cite}
\usepackage{graphicx}  
\usepackage{color}
\usepackage[linesnumbered,ruled]{algorithm2e}
\usepackage{subfig}
\usepackage{tabularx}
\usepackage{rotating}
\usepackage{tikz}
\usepackage{url}
\usepackage{changes}

\ifCLASSINFOpdf
\else
\fi

\begin{document}

\title{Audiovisual Singing Voice Separation}

\author{
        Bochen Li,
        Yuxuan Wang,
        and~Zhiyao Duan
 \thanks{Bochen Li and Zhiyao Duan are with the Department of Electrical and Computer Engineering, University of Rochester, NY, USA. Yuxuan Wang is with the ByteDance Inc. E-mails: bochen1106@gmail.com, wangyuxuan.11@bytedance.com, zhiyao.duan@rochester.edu.}
}

%



\maketitle

\begin{abstract}
Separating a song into vocal and accompaniment components is an active research topic, and recent years witnessed an increased performance from supervised training using deep learning techniques. We propose to apply the visual information corresponding to the singers' vocal activities to further improve the quality of the separated vocal signals. The video frontend model takes the input of mouth movement and fuses it into the feature embeddings of an audio-based separation framework. To facilitate the network to learn audiovisual correlation of singing activities, we add extra vocal signals irrelevant to the mouth movement to the audio mixture during training. 
We create two audiovisual singing performance datasets for training and evaluation, respectively, one curated from audition recordings on the Internet, and the other recorded in house. The proposed method outperforms audio-based methods in terms of separation quality on most test recordings. This advantage is especially pronounced when there are backing vocals in the accompaniment, which poses a great challenge for audio-only methods.
\end{abstract}


\IEEEpeerreviewmaketitle

\section{Introduction}
\label{sec:introduction}

Vocal performance is an important art form of music. The task of singing voice separation is to isolate vocals from the audio mixture, which contains other instrumental sounds that help to define the harmony, rhythm, and genre. 
Singing voice separation is often the first step towards many application-oriented vocal processing tasks including pitch correction, voice beautification, and style transfer, as implemented in some mobile Apps such as WeSing and Smule. It is also often a pre-processing step for other research tasks such as singer identification \cite{berenzweig2002using}, lyrics alignment \cite{fujihara2006automatic}, and tone analysis \cite{fujihara2007music}.

There are various scenarios when video recordings are available for singing performances, such as operas, music videos (MV), and self-recorded singing activities. In pop music, creative visual performances give artists a substantial competitive advantage. Moreover, due to the rapid growth of Internet bandwidth and smartphone users, videos of singing activities are becoming popular in a number of video sharing platforms such as TikTok and Instagram.

Visual information, e.g., lip movement, has been incorporated and shown its benefits in speech signal processing, such as audiovisual speech separation \cite{lu2019audiovisual}, enhancement \cite{afouras2018conversation}, and recognition \cite{petridis2018endtoend}. Visual information has also been incorporated in music analysis \cite{duan2019audiovisual}, such as source association \cite{li2019online,li2017see,li2017audiovisual}, source separation \cite{zhao2019sound}, multi-pitch analysis \cite{dinesh2017visually},  playing technique analysis \cite{li2017videobased}, cross-modal retrieval \cite{li2019query} and generation \cite{chen2017deep,li2018pianist}. For singing performances, however, little work has been done. It is reasonable to think that visual information would also help to analyze singing activities, and in particular, separate singing voices from background music. This is based on the fact that mouth movements and facial expressions of the singer are often correlated with the singing voice signal fluctuations. The advantages of audiovisual analysis over audio-only analysis can be best demonstrated in songs with backing vocals in the accompaniment and songs with multiple singers singing simultaneously. However, to what extent does the incorporation of visual information help singing voice separation is still a question. Different from speech signals, singing voices (except for rap music) generally change slower \cite{mesaros2010recognition}, showing less frequent matching with mouth movements \cite{cadalbert1994singing}. Furthermore, some musically important fluctuations of the singing voice such as pitch modulations show little, if any, correlation with mouth movements \cite{connell2013you}.

Therefore, it is our intention to answer the following research question in this paper: \textit{Can visual information about the singer improve singing voice separation, and if yes, how much?} It is noted that while traditional singing voice separation tasks (e.g., SiSEC\footnote{A community-based signal separation evaluation campaign. \url{https://sisec18.unmix.app/#/}} or MIREX\footnote{Music Information Retrieval Evaluation eXchange. \url{https://www.music-ir.org/mirex/wiki/MIREX_HOME}}) define all vocal components in a song as the singing voice, in this work we define it as separating the solo singing voice from the accompaniments, where the accompaniments may contain backing vocals. We argue that our definition is more musically meaningful as it separates solo, typically presenting the main melody, from accompaniment, typically presenting harmony. Separating the solo voice enables many applications such as pitch refinement and voice beautification for the soloist without affecting the backing vocal sources. The solo singing voice separation problem is somewhat similar to speech enhancement with babble noise. However, music accompaniment is typically much louder and richer in timbre than background noise in speech enhancement settings. In addition, music accompaniment, especially backing vocal, shows very strong correlations with the solo vocal signal. They make the problem at hand very challenging.

To answer the above-mentioned research question, we design an audiovisual neural network model to separate the solo singing voice from the accompaniments that may contain backing vocals. This network model takes both the audio mixture signal and the mouth region of the singing video as input. The audio processing sub-network is designed based on the MMDenseLSTM \cite{takahashi2018mmdenselstm}, the champion of SiSEC2018 (the latest one of SiSEC). The visual processing sub-network uses convolutional and LSTM layers to encode mouth movements of the singer. The audio and visual encodings are fused before they are used to reconstruct the solo singing magnitude spectrogram. The training target of the proposed audiovisual network is to minimize the Mean-Square-Error (MSE) loss of the magnitude spectrogram reconstruction of the solo singing voice. To facilitate the network to learn audiovisual correlation of singing activities, we add extra vocal signals irrelevant to the solo singer to the audio mixture during training.
To investigate the benefits of visual information, we compare the proposed audiovisual model with several state-of-the-art audio-based singing separation methods and an audiovisual speech enhancement method. We further vary the architecture and input of the visual processing sub-network to compare their performances.

One challenge we encounter in this work is the lack of audiovisual datasets of singing. For training, this can be addressed by randomly mixing solo singing videos downloaded from the Internet with irrelevant accompaniment music. We download \textit{a cappella} audition vocal performance videos and randomly mix their audio with accompaniment audio tracks from the MUSDB18 dataset (officially provided by SiSEC2018) to generate  mixtures. We name this the \emph{Audition-RandMix} dataset, and partition it into training, validation and test subsets.
For evaluation on real songs, however, we need audiovisual recordings of singing with its relevant accompaniment music in separate tracks. To our best knowledge, no such dataset exists. Therefore, we record a new audiovisual dataset named \emph{URSing}, where singers are recruited to sing along with prepared accompaniment tracks.

We conduct experiments on both the Audition-RandMix test set and the URSing dataset.
Results on both sets show that the proposed audiovisual method outperforms baseline methods in most test conditions, no matter if the accompaniment tracks contain the backing vocals or not. 
We further conduct subjective evaluations on a cappella video performances in the wild to prove the advantages of our proposed method.

The contributions of this paper include:
\begin{itemize}
    \item The first work to incorporate visual information to the state-of-the-art music source separation framework to address the singing voice separation problem,
    \item A proposal of solo voice separation where backing vocal components, if exist, are regarded as accompaniment tracks, which better fits many application scenarios, and
    \item The first audiovisual singing performance dataset, URSing, free for download\footnote{\url{http://www.ece.rochester.edu/projects/air/projects/URSing.html}}.
\end{itemize}

\section{Related Work}
\label{sec:related}

\subsection{Singing Voice Separation}

Early methods for singing voice separation include non-negative matrix factorization \cite{vembu2005separation}, adaptive Bayesian modeling \cite{ozerov2007adaptation}, robust principal component analysis \cite{huang2012singingvoice}, and auto-correlation \cite{rafii2011simple}. Recently, deep learning based methods are proposed to model convolutional \cite{chandna2017monoaural} or recurrent structures \cite{uhlich2017improving} of magnitude spectral representations of music signals. Some works also learn to reconstruct spectral phases in addition to magnitudes \cite{takahashi2018phasenet,choi2019investigating}, while others directly work on time-domain waveforms with an end-to-end training strategy \cite{lluis2019endtoend,stoller2018waveunet}. A direct comparison of recently proposed methods is available at the SiSEC2018 post. The best performing methods in SiSEC2018 use a DenseNet structure with a recurrent structure to process magnitude spectrograms \cite{takahashi2017multiscale,takahashi2018mmdenselstm}, where the feature reuse strategy inside each dense block greatly reduces the model size. Later some open-sourced methods/tools have been proposed with comparable results, such as Open-Unmix \cite{stoter2019openunmix} and Spleeter \cite{hennequin2019spleeter}. In this paper, we build upon the DenseNet to propose an audiovisual model.

\subsection{Audiovisual Source Separation}
\label{sec:related:audiovisual}

Most audiovisual separation works are proposed for speech signals.
For speech separation, one challenge is the permutation problem where the separated components need to be assigned to the correct talkers. \cite{lu2018listen} specifically address the problem by applying the visual information as a post-processing step to adjust the separation mask. Later the same group proposes to fuse the visual information to an audio-based deep clustering framework to propose an audiovisual deep clustering model for speech separation \cite{lu2019audiovisual}. Another work is described in \cite{ephrat2018looking}, where the input is the mixture spectrogram and the face embeddings of all the appeared speakers in the audio sample. The training target is the complex mask that can be applied to the original spectrogram to recover the complex spectrogram of each speaker. It is noted that speech separation algorithms typically assume a noiseless or less noisy environment in which speech signals are mixed. In addition, speech signals to be separated are typically assumed to be from different speakers. Both assumptions are not true in solo singing separation, as the background music is often quite strong and the backing vocal often comes from the same singer as the soloist.

Speech enhancement aims at separating speech signal from background noise. It is more relevant to singing voice separation from background music considering the foreground-background relations of sources.
\cite{hou2018audiovisual} address the speech enhancement problem using a two-stream structure that takes both noisy speech and frames of the cropped mouth regions as inputs to compute their features. These features are then concatenated by a fusion network which also outputs corresponding clean speech and reconstructed mouth regions. Another audiovisual speech enhancement work proposed in \cite{afouras2018conversation} uses 1D convolutional layers to reconstruct the magnitude spectrogram of the clean speech and uses it to further estimate its phase spectrogram. The input of the visual branch is the feature embeddings on the lip region that are pre-trained on lip reading tasks.

Less work has been proposed for audiovisual music separation. \cite{parekh2017motion} apply non-negative matrix factorization (NMF) to separate string ensembles, where the bowing motions are used to derive additional constraints on the activation of audio dictionary elements. This method, however, is only evaluated on randomly assembled video scenes of string instruments where distinct bowing motions of each player are clearly captured. \cite{zhao2018sound} propose to learn static audiovisual correspondence with cross-modal source localization; The correlation between each pixel in a given video frame and the sound component can be constructed.
Later the same group proposes to learn dynamic audiovisual correspondence \cite{zhao2019sound} which captures correlations between motions and sound fluctuations. Another followup work is to recognize visual gestures for sound source separation \cite{gan2020music}.
This line of research achieves promising results in audiovisual music separation, but have not addressed singing voice separation.

\section{Method}

\subsection{Network Architecture}

\begin{figure}
\centerline{\includegraphics[width=0.9\columnwidth]{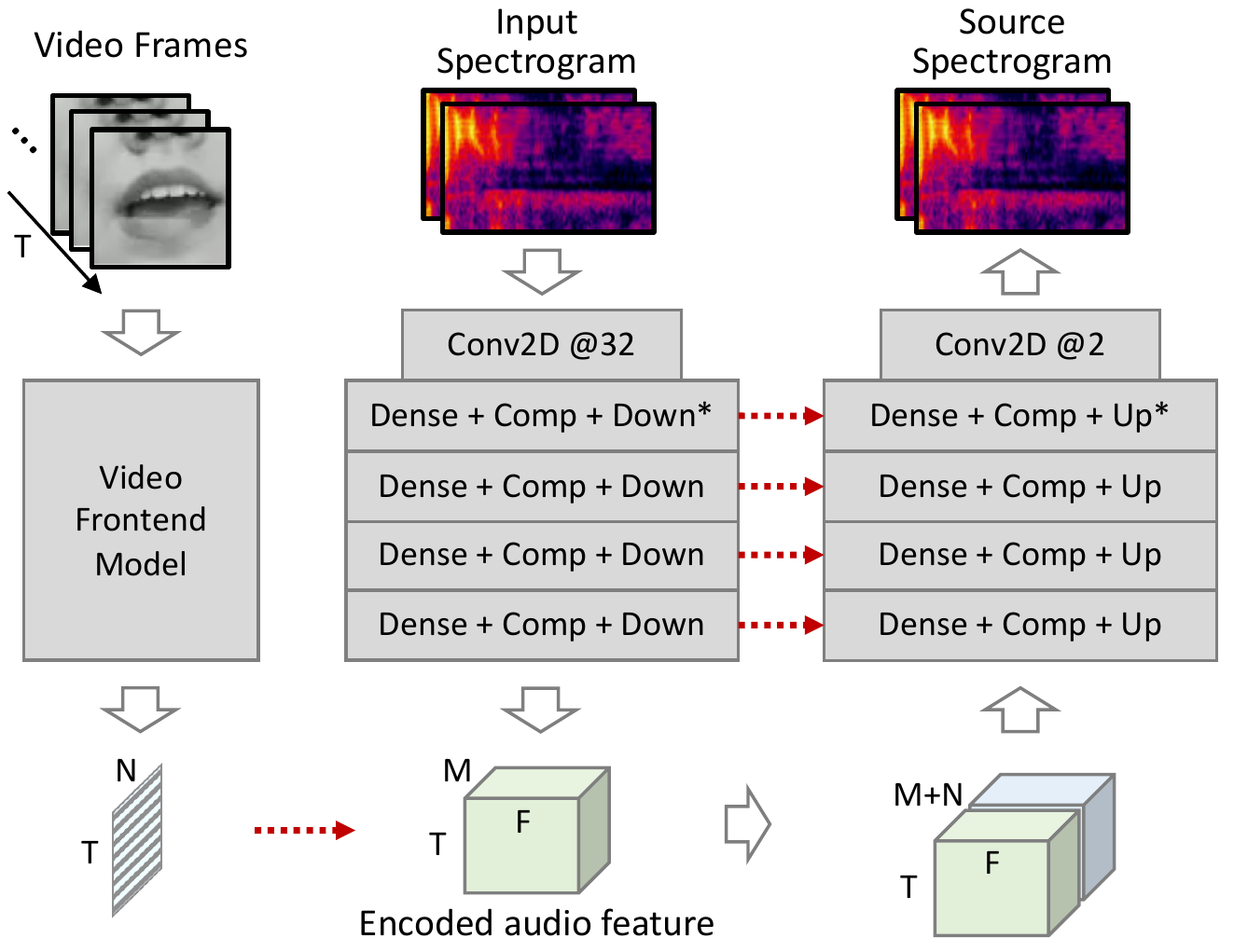}}
\caption{The proposed model structure. Dashed arrows denote the concatenation operation. Downsample/upsample are applied to both time and frequency dimensions in the outer layers (marked by *), while they are only applied to the frequency dimension in the inner layers.
}
\label{fig:network}
\end{figure}

The proposed system builds upon a state-of-the-art audio separation model named MMDenseLSTM \cite{takahashi2018mmdenselstm} with a video front-end model. The MMDenseLSTM model consists of convolutional layers stacked into dense blocks, which alternates downsample/upsample layers to form a multi-scale structure. It first embeds an input magnitude spectrogram into an encoded feature space and decodes it to recover the separated magnitude spectrogram. Skip connections are added as concatenations on the corresponding layers with the same feature map size. This ``encoder-decoder'' structure with skip connections is widely applied in several music separation models \cite{jansson2017singing,stoller2018waveunet,zhao2019sound,liu2018denoising}. The video front-end model extracts visual features from mouth movements, which are fused with the encoded audio feature. The network structure is illustrated in Figure \ref{fig:network}. We explain each part of the model in detail as follows.

\subsubsection{Audio Separation Model}

Following MMDenseLSTM, our audio separation model consists of:

\begin{itemize}

\item Dense Block. It applies 2D convolutional layers\footnote{A convolutional layer includes BatchNormalization+ReLU+Conv2D throughout the paper.} and the output feature maps of all layers are concatenated with each other along the channel dimension. This structure reuses the feature maps from previous layers and greatly reduces the model size.

\item Compression layer. It is a convolutional layer with 1$\times$1 kernels. We use a compression ratio of 0.2, which means that the number of feature maps (channels) is reduced by 80\% after each compression layer.
We apply a compression layer right after each dense block, which improves the model compactness.

\item Downsample-Upsample. 
These layers are used to resize the feature maps without changing the the number of channels. 
Downsample layers are average pooling with 2$\times$2 kernels after the first compression layer, and 1$\times$2 kernels in the following layers. 
In other words, downsampling is performed along both the time and freuqncy dimensions in the first layer, but only to the frequency dimension in other layers.  
Symmetrically, upsample layers apply transposed convolutional layers with 2$\times$2 kernels and strides at the last upsample layer but 1$\times$2 for the other layers. 
Different from \cite{takahashi2018mmdenselstm} where downsample/upsample always addresses both time and frequency dimensions in multiple scales, our proposed strategy downsamples/upsamples the time dimension only once, making the audio stream have the same frame rate as the video stream. The encoded audio spectrogram feature is denoted as $\mathbf{S}_A \in \mathbb{R}^{M \times T \times F}$, with the channel ($M$), downsampled time ($T$), and frequency ($F$) dimensions.

\item Multi-Band. Following \cite{takahashi2017multiscale}, we also equally divide the spectrogram into a low-frequency band and a high-frequency band and apply the above-mentioned encoder-decoder structure on each sub-band. 
The dense blocks of low-frequency band have a higher channel number. Detailed parameters can be referred to \cite{takahashi2017multiscale}.

\end{itemize}

\subsubsection{Video Front-End Model}

We propose to apply a separate input branch to parse the input video stream and fuse it with the encoded audio features. The video stream is a sequence of mouth region images in consecutive video frames. Raw RGB values are normalized to zero mean and unit variance. 
We use 2D convolutional layers and LSTM layers to extract the visual features from the input RGB frames. Each frame is processed independently sharing the same CNN parameters before being fed into the following LSTM layer. Detailed network architecture is Conv2D@16 (channel number is 16), Conv2D@16, Conv2D@32, Conv2D@32, FC@256, LSTM@128, and FC@$N$, where $N$ is the dimension of the encoded feature vector for each video frame. The input video stream with $T$ frames results in a feature map $\mathbf{S}_V \in \mathbb{R}^{N \times T}$. There is no pooling operation along the time dimension thus the temporal information is preserved.

\subsubsection{Audiovisual Fusion}

The extracted visual feature map from the video branch is fused with the encoded audio spectrogram feature map $\mathbf{S}_A$. To do so, the visual feature map $\mathbf{S}_V$ is inflated along the third dimension and then concatenated with the audio feature to obtain the audiovisual feature $\mathbf{S}_{AV} \in \mathbb{R}^{L \times T \times F} $, where $L = M + N$ is the concatenated channel dimension. Note that the temporal information from both the audio and video branches is correlated during this fusion; This is different from some works where audiovisual fusion is performed on feature maps that aggregate information along time.

In addition to minor structural changes, we also drop the LSTM structure of the original MMDenseLSTM model \cite{takahashi2018mmdenselstm} when we design the audio branch of our proposed model. This follows the observation that the addition of the LSTM structure does not achieve substantial improvement in SiSEC2018 yet the number of parameters would be increased significantly for audiovisual fusion.

\subsection{Training}

We train the model to predict the magnitude spectrogram of the source signal and use the original mixture's phase to recover the time-domain waveform.
Many spectral-domain source separation methods, especially those for speech signals, use a spectrogram mask as the training target; This mask is then multiplied element-wise with the mixture signal's magnitude spectrogram to recover the source magnitude spectrogram. For music separation, some recent works train networks to directly output the source magnitude spectrogram \cite{uhlich2017improving,takahashi2018mmdenselstm} using a Mean-Squared-Error (MSE) loss. We follow the same way and take the source magnitude spectrogram as the training target. However, we have a mask layer that regularizes the feature maps into the range of [0, 1] using a Sigmoid function and multiplies the mask layer with the input spectrogram. We find that this is necessary for fusing the visual input into the audio feature, as the audiovisual method even degrades the separation performance in some experimental settings when the mask layer is not in place. We have a comparative 
experiment in Section \ref{sec:experiments:video_models}.

The model input is the magnitude spectrogram of the original audio mixture which contains both the solo vocal and background music, and the mouth region of the video frames corresponding to the vocals. The output is the magnitude spectrogram of the source audio. 
Note that the magnitude spectrogram has been converted to logarithm scale followed by normalization along each frequency axis, which better weighs the contribution of high frequency bins.
Compared to the audio mixture input, the visual input provides much less information about the source signals, therefore, the training loss may not be propagated back sufficiently into the visual branch, making the audiovisual network difficult to train. One way to address this is to explicitly learn audiovisual matching, either through pre-training \cite{lu2018listen} or early audiovisual fusion \cite{lu2019audiovisual}. Another way might be to add visual reconstruction as another training target, leading to a chimera-like network structure \cite{hou2018audiovisual}.

In this work, we address this problem by adding some extra vocal components to the original mixture, which are not related to the mouth movements and thus are not included in the target vocal spectrogram. This is similar to adding an additional speaker in the training data in the case of audio-visual speech separation \cite{ephrat2018looking}, which forces the model to learn audiovisual correlations after the fusion and only separate the vocal components that are related to the visual input.
Note that in the training samples all of the vocal and accompaniment components are randomly mixed, so neither the extra vocal components or the solo vocal components have harmonic relations with the accompaniment tracks. In the experiments, we show that the strategy of training with randomly generated vocal-accompaniment pairs performs decently on real songs.

\section{Dataset}
\label{sec:dataset}

Since there is no publicly available audiovisual singing voice dataset containing isolated vocal tracks, we collect our own data for training and evaluating the proposed method.

\subsection{A Cappella Audition Vocals (AAV)}

We curated 491 YouTube videos of solo singing performances by querying the YouTube search API with the keyword ``Academic Acappella Audition''. We only selected video excerpts where the singer faces the camera and sings without accompaniment.
The total length of these excerpts is about 8 hours. As it is difficult to find relevant and appropriate accompaniment tracks, in our experiments we simply randomly chose instrumental accompaniment tracks (from the ``accompaniments'' track in the MUSDB18 dataset) and mixed them with the solo singing excerpts to create singing-accompaniment mixtures.

The randomly mixed samples are used for training, validation, and evaluation. Before the mixing process, vocals in AAV are divided into training/validation/evaluation sets roughly as 8:1:1 (50 tracks for evaluation). Accompaniment tracks from MUSDB18 (which contains a wide range of music genres and instrument types) are also divided into the three sets following the official way (also 50 tracks for evaluation). Then mixing is applied on each split independently to form the training/validation/evaluation sets. Volume of each track is normalized using the root-mean-square (RMS) value. For training and validation sets, each track is split into short samples (around 2.5 seconds) for random mixing, resulting in a massive amount of mixed samples. We do not balance the volume of each individual sample so the mixing may have different SNRs. For evaluation, mixing is performed on a random bijection between the 50 vocals and 50 accompaniments. For each mixing, we pick a 30-second excerpt (with both vocal and accompaniments present) for evaluation, following the same strategy as the MUSDB18 dataset. This set is referred to as ``Audition-RandMix'' in the following experiments. In addition, we randomly add extra vocals from another vocal track in the test set (with the same RMS value as the target vocal) into these mixings for evaluations, referred as ``Audition-RandMix (v+)'', in order to explore the model performance in more challenging cases. Note that all the testing samples in this condition are not musically meaningful and cannot represent real songs.

\subsection{URSing}

To evaluate the proposed method in more realistic singing performances, we create the University of Rochester Multi-Modal Singing Performance Dataset (URSing). In this paper, we only use the URSing dataset for evaluation. A brief description of the creation process is described below.

\begin{figure}
\centerline{\includegraphics[width=\columnwidth]{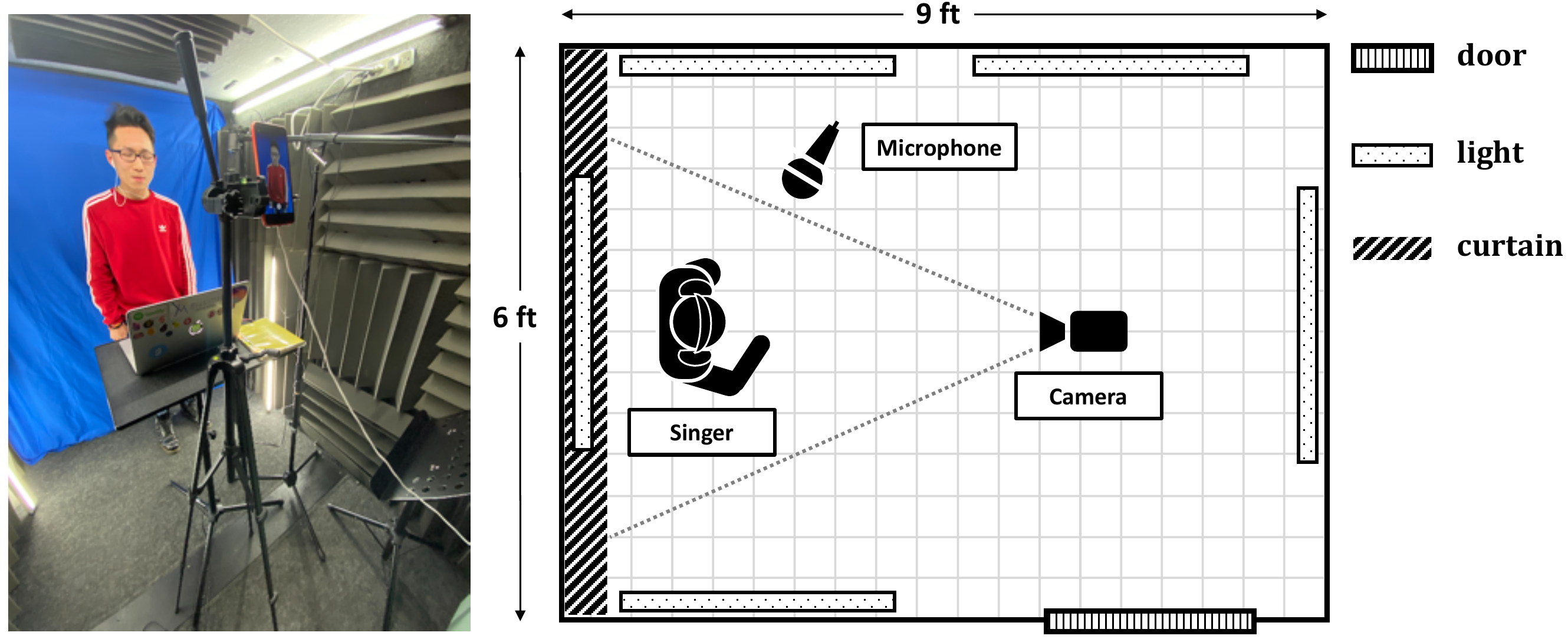}}
\caption{A sample photo and floor plan of the sound booth for the recording process of the URSing dataset.}
\label{fig:floor_plan}
\end{figure}

\subsubsection{Singer Recruiting} Singers are students at the University of Rochester. Audition is performed to filter out unqualified singers who could not sing in tune. Each participant receives \$5 for recording each song, and is allowed to record up to 5 songs.

\subsubsection{Piece Selection}
To ensure high recording efficiency, the singers pick their own songs and their favorite accompaniment tracks to sing along. We do not put constraints on song genres, but filter out songs of which the accompaniment tracks are of low sound quality.

\subsubsection{Recording} To ensure synchronization, the singers listen to the accompaniment track through earphones while recording their singing voice. Their voices are recorded using an AT2020 condenser microphone hosted by Logic Pro X, and their videos are recorded using iPhone 11. The recording is conducted in a semi-anechoic sound booth. A sample photo and the floor plan of the sound booth are shown in Figure \ref{fig:floor_plan}.
    
\subsubsection{Post-processing} For each solo vocal recording we use the following plug-ins to simulate the typical audio production procedure in commercial recordings: a) static noise reduction (\textit{Klevgrand Brusfri} and \textit{Waves X-noise}), b) pitch refinement (\textit{Melodyne}), c) sound compression (\textit{Fabfilter Pro-C 2}), and d) reverberation (\textit{Fabfilter Pro-R}).
We also adjust the vocal volume to balance it with the accompaniment track. Beyond this, we do not perform any other editing on the audio recording (e.g., time warping or rhythmic refinement) to preserve the synchronization with the visual performance. To synchronize the audio recording captured by the AT2020 microphone with the video recording captured by the smartphone, we use the audio recording captured by the built-in microphone of the smartphone as the bridge, through cross correlation.

\subsubsection{Annotation}
Since the mouth movements are mostly relevant to the singing performance, we provide the annotations of the mouth regions in the dataset. This is performed using the Dlib library \cite{king2009dlibml}, an automatic tool for facial landmark detection, followed by manual check. The mouth region is represented as a square bounding box with the side length equal to 1.2 times of the maximum horizontal distance for all mouth landmarks.

\begin{figure}
\centerline{\includegraphics[width=0.8\columnwidth]{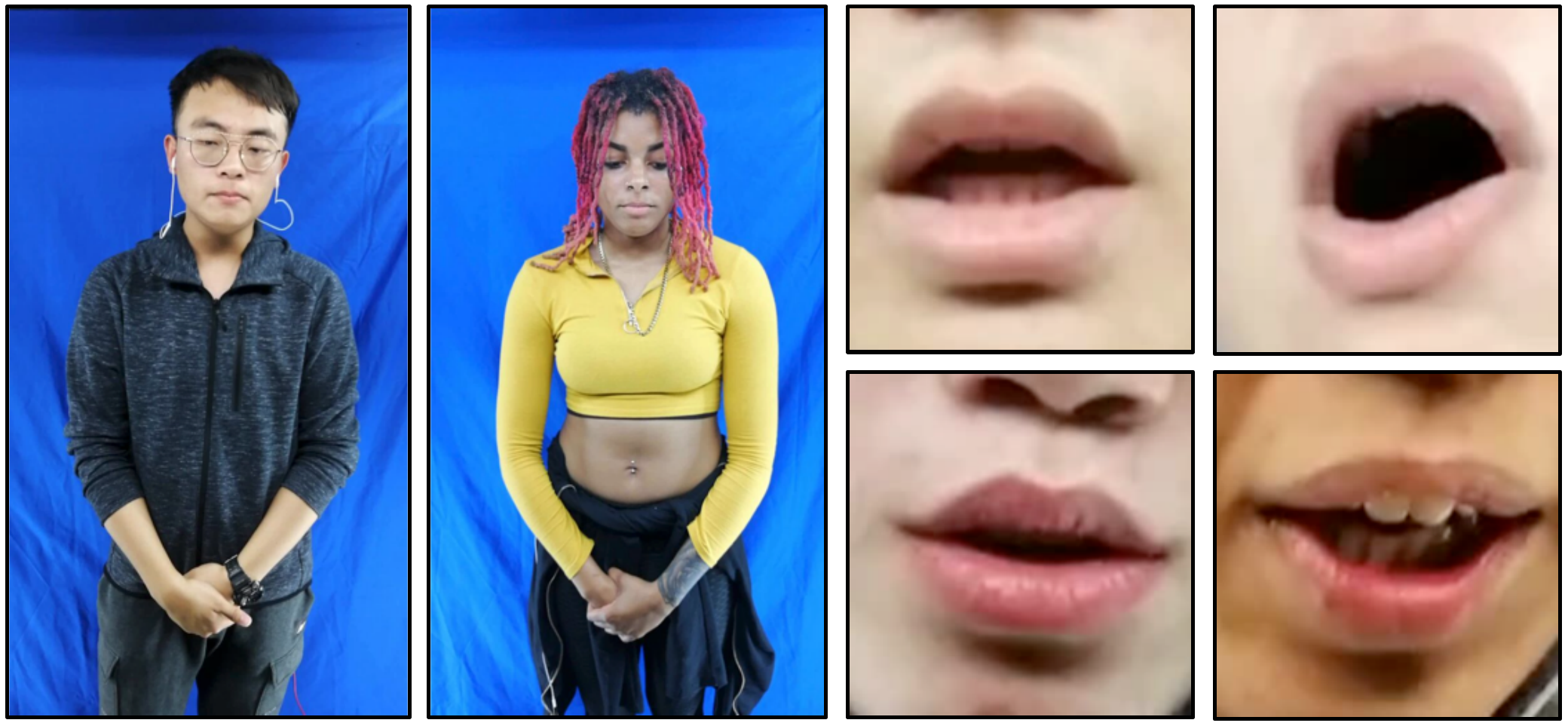}}
\caption{Examples of video frames of the URSing dataset and cropped mouth region pictures as the input to the video branch of the proposed method.}
\label{fig:dataset_sample}
\end{figure}

This results in 65 songs, totaling 4 hours of audiovisual recordings of singing performance. For each song, we provide the audio recording of the solo singing voice (in WAV, 44.1 KHz, 16 bits, mono) and the corresponding accompaniment audio track (same format, mono or stereo). We also provide the video recording of the soloist (in MP4, 1080P portrait, 29.97 FPS), where the soundtrack is the mixture (direct sum) of the solo vocal and the accompaniment tracks. Note that when we prepare the accompaniment tracks, we do not avoid the tracks containing backing vocals, as they are the challenging and useful cases to study in this paper.
Example video frames and cropped mouth region pictures are provided in Figure \ref{fig:dataset_sample}.

We also choose a set of 30-sec excerpts where both solo vocal and accompaniment tracks are prominent to form a benchmark evaluation set. Specifically, for each of the 65 songs, we choose one 30-sec excerpt without backing vocal and one with back vocal, if such excerpts are available. We provide this information in the metadata. 
This results in 54 excerpts with accompaniment tracks that only contain instrumental components (referred as ``URSing'' in the following experiments) and 26 excerpts with accompaniment tracks that also contain backing vocals (referred as ``URSing (v+)''. The latter, presumably, are more challenging for solo vocal separation and more useful for showing advantages of audiovisual methods. In this paper, since we do not use any songs from URSing for training, we only use these 30-sec excerpts for evaluation.


\section{Experiments}

\subsection{Implementation Details}
For audiovisual singing videos, audio is downsampled to 32 KHz. We use a frame length of 1024 and a hop size of 640 (20 ms) for spectrogram calculation. Video data is converted to 25 FPS (equivalent to 40 ms frame hop size), and the frame size of mouth regions is interpolated into 64 $\times$ 64. Each data sample is 2.56 seconds long, containing 128 audio frames and 64 video frames. The input/output audio spectrogram has the shape of 2$\times$128$\times$513 (channels $\times$ frames $\times$ frequency bins), and each input video stream has the shape of 64$\times$64$\times$64 (frames $\times$ width $\times$ height).

During training, for half of the training/validation samples, we add extra vocal components that are not related to the mouth movements to encourage the model to learn audiovisual correlations. 
We use batch size of 8 for training on a TITAN X GPU with 11.9 GB graphic memory. It takes about 40 hours to train for 50 epochs. We adopt early stopping when the validation loss does not decrease for 10 consecutive epochs.

For evaluations, we calculate the signal-to-distortion ratio (SDR) between the separated vocal waveforms and the ground-truth ones using the BSS Eval toolbox V4, same as the evaluation measure applied in SiSEC2018. Specifically, for each 30-sec evaluation excerpts, we calculate the median SDR over all 1-sec audio segments.

\subsection{Baselines}

We first use the original mixture recording (referred as ``MIX'' in the experiments) as the separated vocal for evaluation on our dataset. This sets lower bounds of separation results without any separation techniques. Then we apply two oracle filtering techniques that utilize ground-truth source signals: The ideal binary mask (IBM) assigns each time-frequency bin to the predominant source.
The ideal ratio mask (IRM) distributes the power of each time-frequency bin into different sources according to the power ratio of the ground-truth sources. The IBM and IRM set upper bounds for time-frequency masking-based source separation methods.

We then compare our proposed method with several audio-based music separation methods as baselines. 
\begin{itemize}
    \item RX7. A commercial software developed by iZotope\footnote{\url{https://www.izotope.com}}. We apply batch processing of the ``music rebalance'' function with the preset ``isolate vocals'' on ``medium'' level. Training data for the model inside this software is unknown to us.
    \item UMX \cite{stoter2019openunmix}. An open-sourced separation tool known as ``Open-unmix'' . The model employs the BLSTM structure and is trained on the MUSDB18 dataset.
    \item Spleeter \cite{hennequin2019spleeter}. An open-sourced separation tool  with a CNN+Unet model trained on their in-house dataset of 24,097 songs. It achieved best separation results among all open-source tools on the evaluation set of the MUSDB18 dataset up to date. 
    \item Spleeter-train. Same model as ``Spleeter'' but trained on our Audition-RandMix dataset using the same conditions as those for our proposed audiovisual method as a direct comparison.
    \item MMDenseLSTM \cite{takahashi2018mmdenselstm}. The method that achieved the best results in SiSEC2018, even without training on extra data. We implemented this method from scratch. Our implementation has been validated by achieving similar results to the reported ones on MUSDB18, following SiSEC2018's official train/test split. We then trained this model on our Audition-RandMix dataset as a direct comparison.
\end{itemize}


We also implement an audiovisual speech enhancement method named AVDCNN proposed in \cite{hou2018audiovisual}. This method applies 2D CNNs to take noisy speech and the mouth region visual recording as inputs, fuses encoded audio and visual features to output the enhanced speech signal as well as reconstructed video frames of mouth movements. After the fusion layers, we used LSTM instead of fully-connected layers as used in \cite{hou2018audiovisual}, which shows higher performance in our experiment scenarios.

We choose audiovisual speech enhancement instead of audiovisual speech separation as the baseline, because we believe that speech enhancement is more relevant to singing voice separation from background music in terms of foreground-background relations of sources, as explained in Section \ref{sec:related:audiovisual}. In addition, audiovisual speech separation usually assumes the availability of all talkers, while in our setting, only the video of the solo singing voice is used.

We present the model sizes of baseline models that are open-source or implemented by us in Table 1, together with that of the proposed model.

\begin{table}[]
\label{tb:model_size}
    \scriptsize
    \centering
    \begin{tabular}{ c|ccccc } 
     Method & UMX & Spleeter & MMDenseLSTM & AVDCNN & Proposed \\ 
     \hline
     Parameter \\ ($\times 10^6$)
     & 8.5 & 19.7 & 1.22 & 11.3 & 2.05 \\ 
    \end{tabular}
    \caption{Comparison of model size of different methods.}
\end{table}

\subsection{Objective Evaluation on Synthetic Mixtures}

\begin{figure*}
\centerline{\includegraphics[width=1.5\columnwidth]{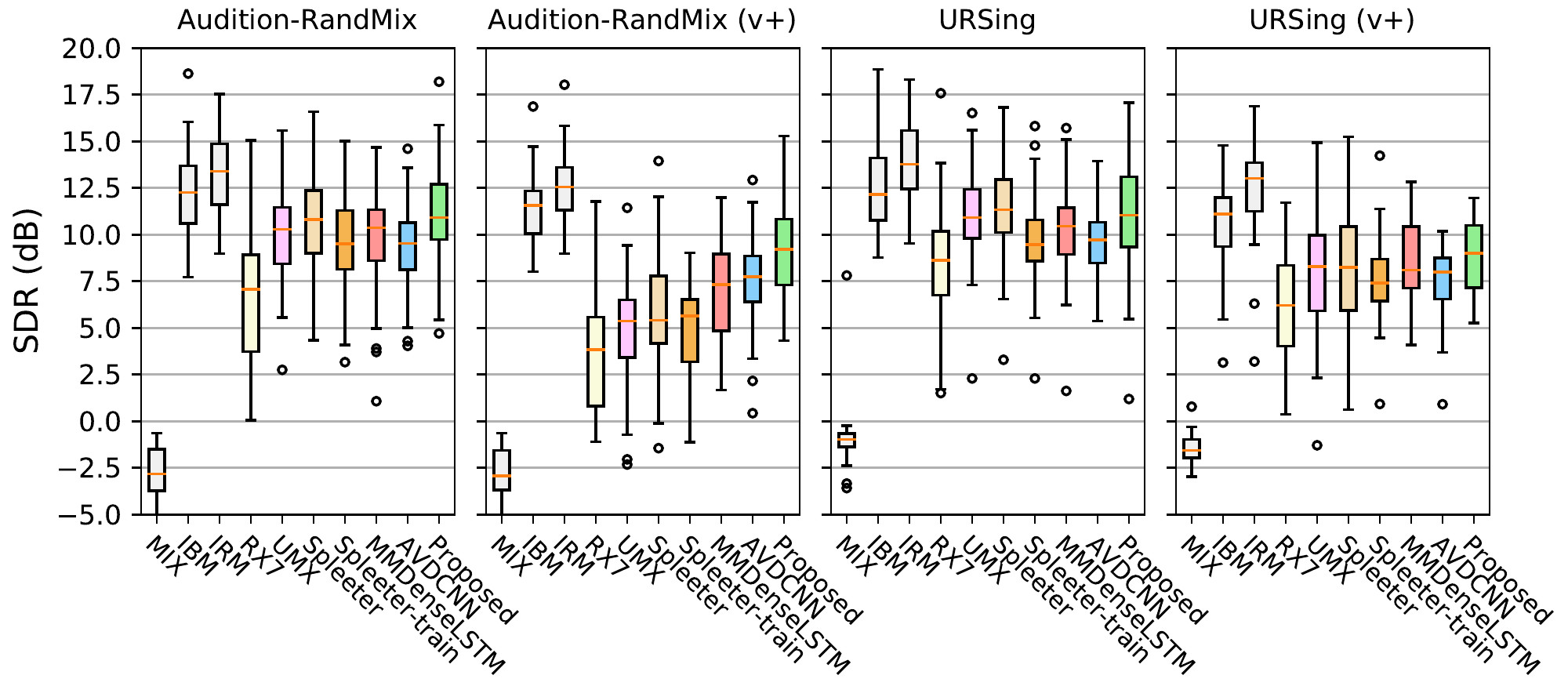}}
\caption{The SDR (dB) comparison on separated solo vocals with different methods on different evaluation sets. (``v+'' denotes for songs where accompaniments contain vocal components.)} 

\label{fig:exp_overall_sdr}
\end{figure*}



We evaluate the comparison methods on the four test sets described in Section \ref{sec:dataset}: Audition-RandMix, Audition-RandMix (v+), URSing, and URSing (v+). ``v+'' means that the accompaniments contain vocal components. Note that all these songs are synthetic mixtures, e.g., Audition-RandMix is random mixed samples and URSing is recorded in controlled environment.
Boxplots of SDR results are shown in Figures \ref{fig:exp_overall_sdr}, where each data point in the boxplots is the median SDR of the separated vocal of all 1-sec segments of a 30-sec excerpt. The horizontal line inside each box indicates the median value across all excerpts. Several interesting observations can be made from the results.

\subsubsection{Benefits of Visual Information}

The proposed method outperforms all audio-based separation baselines in most of the evaluation sets.  This shows the advantage of incorporating visual information about the singer's mouth movement for solo singing voice separation. 
Among the audio-based baseline methods, MMDenseLSTM is much stronger than RX7, because MMDenseLSTM is our own implementation and is trained on our dataset while RX7 is not. However, Spleeter slightly outperforms our proposed system on the URSing set. We believe that this is because Spleeter is trained on a much larger in-house dataset that contains 24,097 songs totalling 79 hours. This is verified by the fact that, Spleeter-train, the same model as Spleeter but trained on our dataset as a fair comparison, does not outperform MMDenseLSTM nor the proposed method. We suggest that this is because our proposed model (and MMDenseLSTM) has a much smaller model size than Spleeter, making it less prone to overfitting given a small training set.

Comparing songs with backing vocals (Audition-RandMix (v+) and URSing (v+)) to songs without backing vocals (Audition-RandMix and URSing), we can see that the outperformance of the proposed method is better pronounced on songs with backing vocals. Wilconxon signed-rank tests show that the improvement of the proposed method over MMDenseLSTM on Audition-RandMix (v+) and URSing (v+) are both significant, with $p$ values of $6.2 \times 10^{-3}$ and $4.3 \times 10^{-2}$, respectively. We argue that this is because audio-only methods tend to assign all the vocal components to the separated singing voice, while the proposed audiovisual method learns to only separate the vocal signals that are correlated to the solo singer's mouth movements.

The reason that the improvement is more pronounced on Audition-RandMix (v+) than on URSing (v+), we argue, are twofold: 1) backing vocals in URSing (v+) are not as strong as the intentionally added backing vocals in Audition-RandMix (v+), and 2) backing vocals in URSing (v+) often overlap with solo vocals and share the same lyrics, showing high correlations with the mouth movements of the solo singer, while the added backing vocals in Audition-RandMix (v+) are irrelevant to the solo vocal.

\begin{figure}
\centerline{\includegraphics[width=0.9\columnwidth]{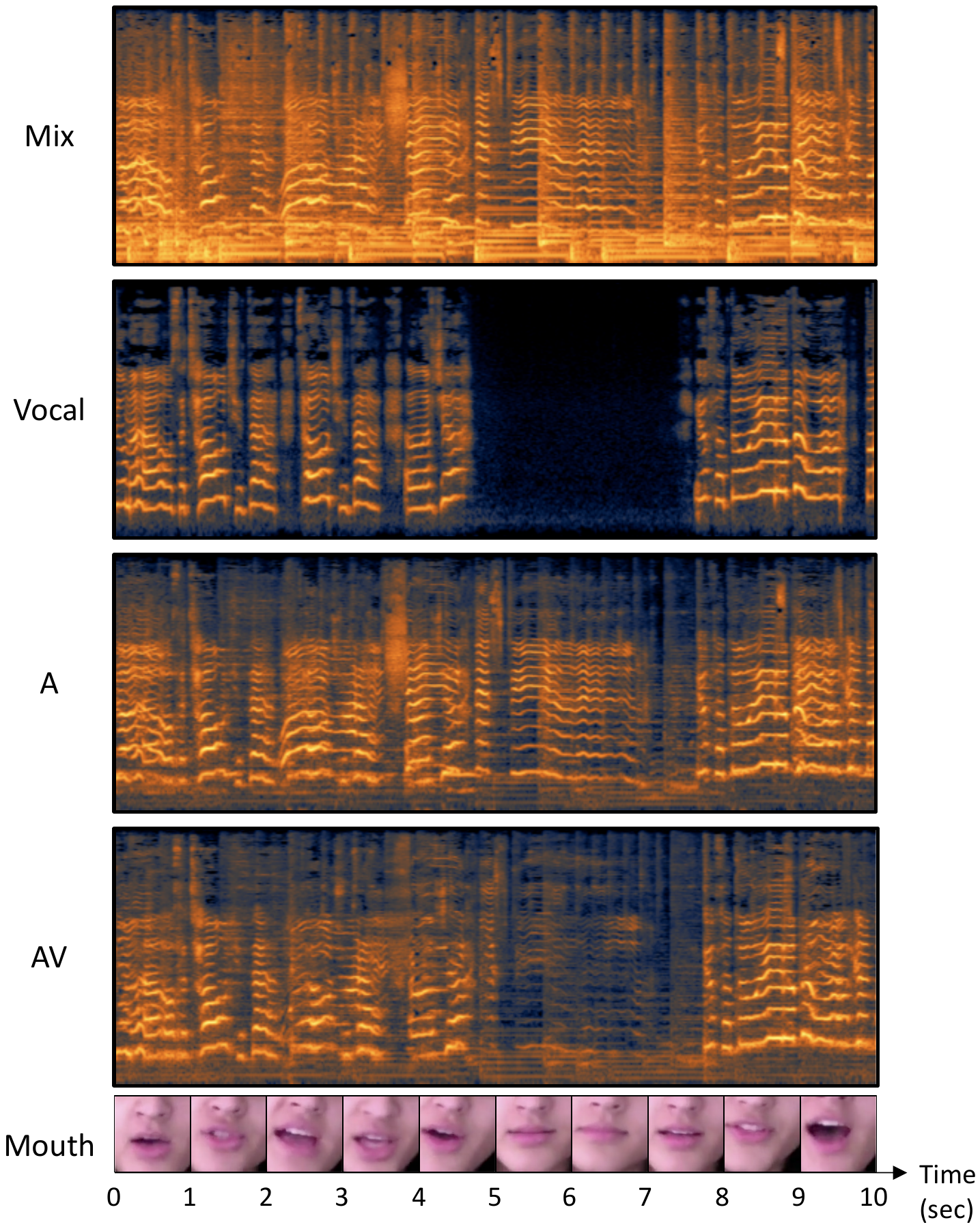}}
\caption{One 10-sec example comparing audio-based separation (MMDenseLSTM) with audiovisual separation (proposed) on a song excerpt with strong backing vocals. The four spectrograms from top to bottom are original mixture, ground-truth vocal, audio-based vocal separation, and audiovisual vocal separation. This sample result has 10-sec long, and one mouth frame of each second is attached.)}
\label{fig:observation}
\end{figure}

Figure \ref{fig:observation} shows one 10-sec sample as an extreme case to compare the spectrograms of audio-based MMDenseLSTM method and the proposed audiovisual method when backing vocal components are strong (e.g., the middle part of the sample). We also show the mouth movement in several frames throughout this excerpt. It can be seen that MMDenseLSTM recognizes the backing vocal components in the middle frames as the solo vocal, while the audiovisual method suppresses those components significantly.

On songs without backing vocals, the outperformance of the proposed method can still be observed.
Subjective listening by the authors suggests that the visual information helps to reduce high-frequency percussive sounds from the solo vocal, as the former do not correlate with mouth movements well.

\subsubsection{Superiority of Proposed Audiovisual Architecture}

The proposed method outperforms the audiovisual speech enhancement baseline significantly in all evaluation sets. Note that the baseline is trained and evaluated on the same dataset as the proposed method. This shows the superiority of the proposed network architecture on the solo singing voice separation task. In particular, we argue two main reasons. First, the proposed model utilizes the commonly used U-net structure with skip connections, which generally achieves good results in music separation \cite{jansson2017singing,stoller2018waveunet,takahashi2017multiscale}. Second, in our audiovisual fusion scheme we preserve the temporal correspondence, which prevents a substantial increase of the number of trainable parameters in the fusion layer. This is important when the DenseNet-based audio sub-network has a small model size. The variations of different video sub-networks, however, does not make much difference on the separation performance, as we analyzed in Section \ref{sec:experiments:video_models}. 

\subsubsection{Limitations and Room for Improvement}

Compared with reported SDR values in SiSEC2018, the SDR values in Figure \ref{fig:exp_overall_sdr} are much higher. For example, MMDenseLSTM reaches over 10 dB on URSing but only less than 7 dB in SiSEC2018 (method ``TAK1'' in \cite{stoter2018signal}). We argue that the songs used in SiSEC2018 (i.e., the MUSDB18 dataset) are professionally recorded, mastered and mixed vocals. They often contain complex components such as polyphonic vocals, background humming, and strong reverberation. They are mastered and mixed by professional music producers to intentionally make them better fused into the background music. In contrast, the ground-truth vocals in our datasets are solo vocals recorded in controlled environments with limited vocal effects added. 
It is reasonable to believe that the benefits of visual information can be further demonstrated on more professionally produced songs. In addition, the performance difference between the Audition-RandMix test sets and the URSing test sets seems to be small for all methods, including the oracle results. This shows that randomly mixed songs, although lacking harmonic and rhythmic coherence, are not easier to separate than the more realistically mixed songs, suggesting that it may be reasonable to use randomly mixed songs to train the methods \cite{luo2017deep}. However, whether this is still true for professionally produced songs is still a question.


On the other hand, there is still some gap between the proposed method and the oracle results on the SDR metric in our evaluation sets. It is likely that this gap will be even bigger on professionally produced songs. This suggests that much work can be done to improve the separation performance. For example, time-domain separation for the audio branch may further improve the performance \cite{luo2018tasnet}.



\subsection{Subjective Evaluation on Professional A Cappella Songs}

\begin{figure}
\centerline{\includegraphics[width=0.9\columnwidth]{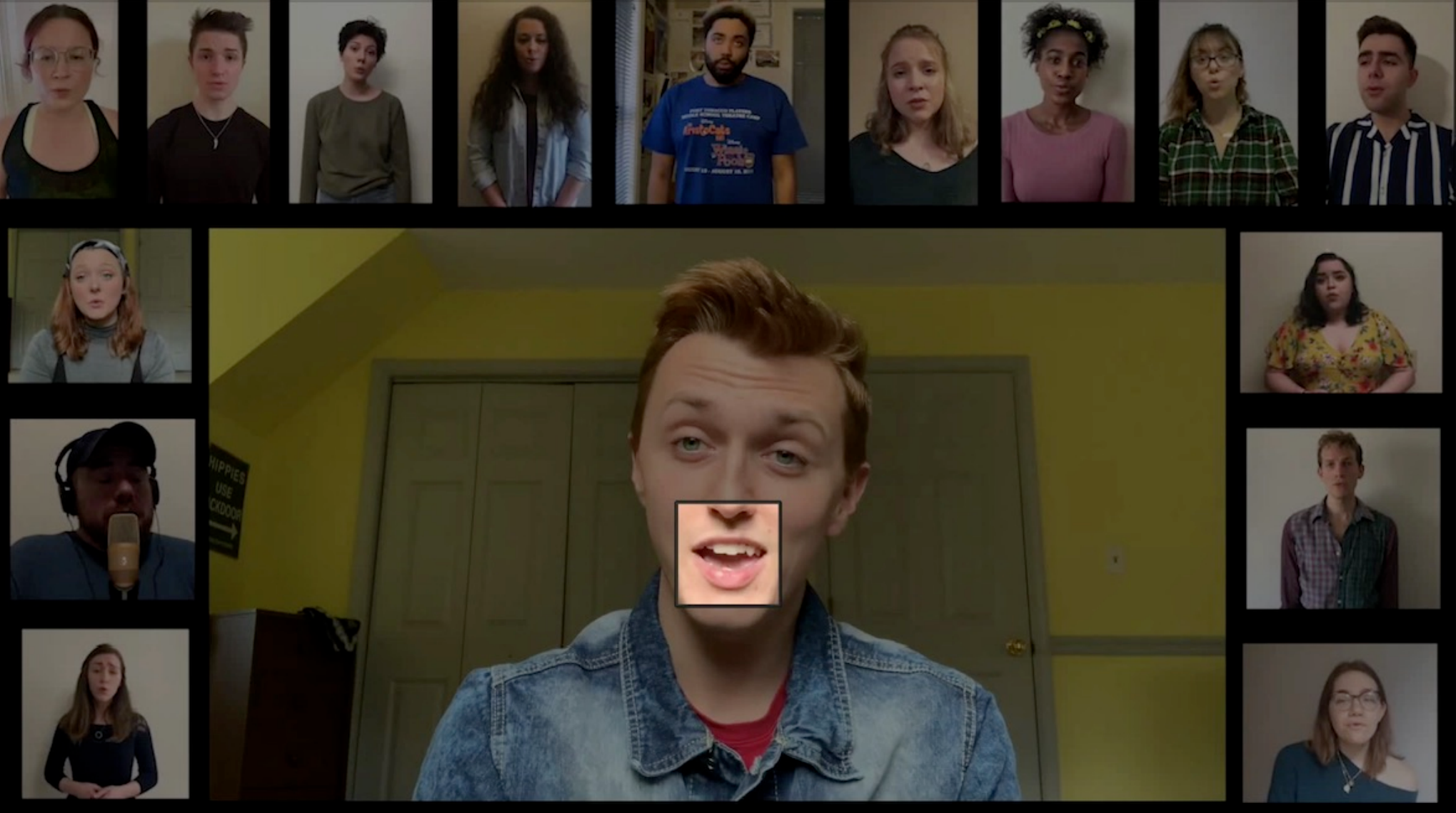}}
\caption{One sample frame of an a cappella song for subjective evaluation.}
\label{fig:example_song}
\end{figure}

\begin{figure}
\centerline{\includegraphics[width=\columnwidth]{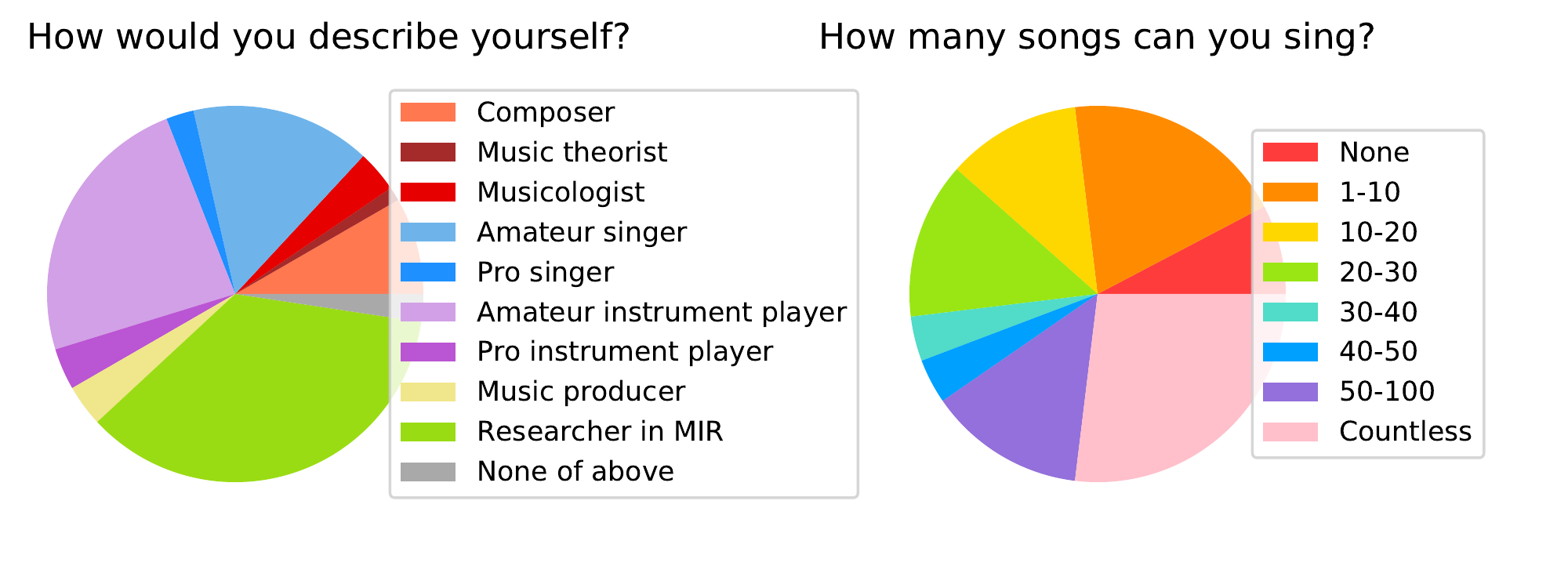}}
\caption{Statistics of the 26 subjects' musical background related to the subjective evaluation.}
\label{fig:exp_subjective_statistics}
\end{figure}

\begin{figure}
\centerline{\includegraphics[width=\columnwidth]{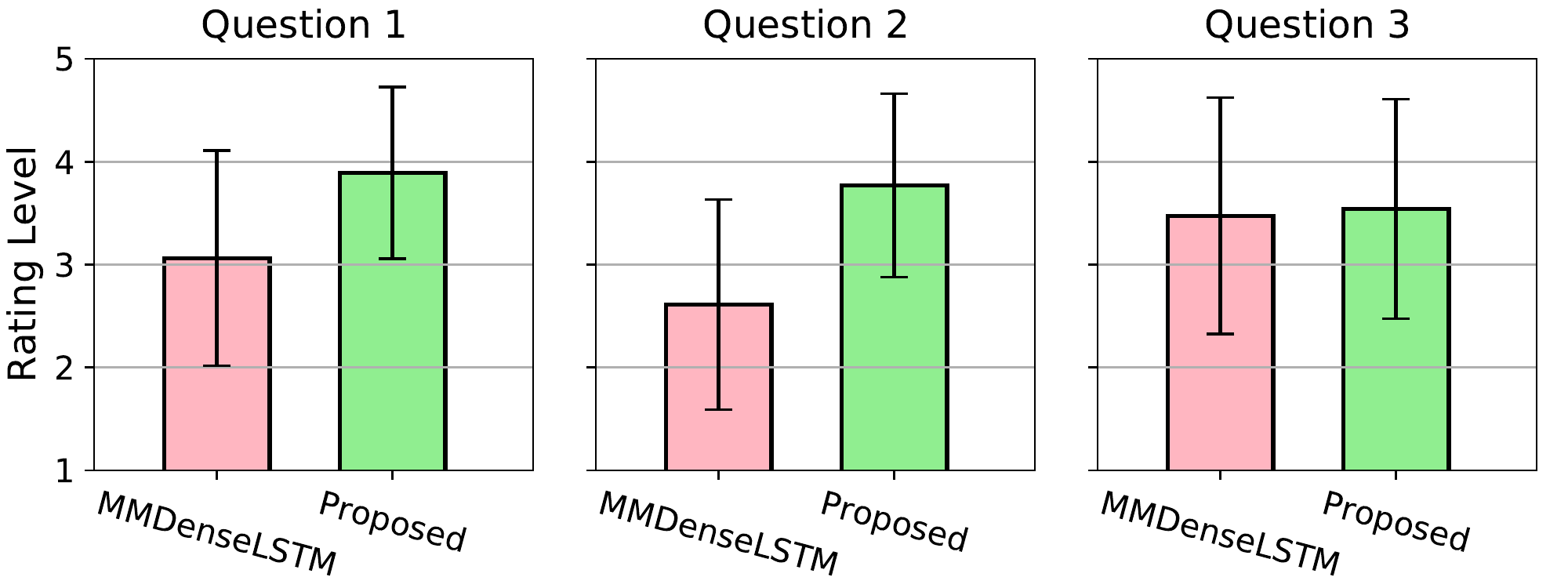}}
\caption{The subjective ratings of the separation quality in response to the three questions. Each error bar shows mean $\pm$ standard deviation.}
\label{fig:exp_subjective_evaluation}
\end{figure}

In this section, We further evaluate the benefits of visual information incorporated in our proposed method on real a cappella songs in the wild. We collect 35 audiovisual a cappella recordings from YouTube. These collections represent the extreme cases where all the accompaniment components are vocals (except for several cases where additional percussive instruments are also present), to study how much the proposed audiovisual method is advantageous while the audio-based method is very likely to fail. Here we use the MMDenseLSTM baseline as the audio-based method for comparison, which yields the best separation results among audio-based baselines. Most of these songs are chorus performance with a solo singer accompanied by harmonic vocals and/or vocal beatbox, while some are performance with multiple solo singers. We only keep the videos where the solo singer's mouth is visible and clear, without video shot transition for at least 10 seconds. A sample frame of one song is shown in Figure \ref{fig:example_song} with the mouth region of the targeted solo singer highlighted.

As we do not have access to the source tracks, we cannot evaluate the separation performance using common objective evaluation metrics. Instead, we conduct a subjective evaluation over 51 people. 
Some subjects are students or faculty from the University of Rochester, others are subscribers from the International Society for Music Information Retrieval (ISMIR) community. Statistics of the subjects' music background is shown in Figure \ref{fig:exp_subjective_statistics}. Each survey asks a subject to rate 7 of the 35 songs, and each subject may take more than one surveys. For each song, the subjects first watch a 10-sec excerpt of the original performance and then watch the same video twice with the solo singing voice separated by two different singing voice separation methods in a random order to rate the separation quality. Due to the variations across these songs, the original recording serves as a reference for a consistent scoring scheme. For each video we also highlight the mouth region of the target solo singer (see Figure \ref{fig:example_song}) to help subjects focus on the corresponding solo voice. The specific evaluation questions are:
\begin{itemize}
\item Question 1: \textit{What do you think about the overall separation quality for the targeted singer?}
\item Question 2: \textit{What do you think about the separation quality in terms of removing backing vocal accompaniments in the separated solo voice?}
\item Question 3: \textit{What do you think about the separation quality in terms of not introducing artifacts into the separated solo voice?}
\end{itemize}
The subjects need to answer each question using a scale from 1 to 5, where ``1'' represents \textit{Very bad} and ``5'' represents \textit{Very good}. The three questions are related to the common definitions of the three objective source separation evaluation metrics, SDR, SIR, and SAR, respectively.

The results of the subjective evaluations are presented in Figure \ref{fig:exp_subjective_evaluation}. According to the collected responses for Question 1, the proposed audiovisual method is rated significant higher than the baseline audio-based method (Wilconxon signed-rank test shows a $p$ value of $3.5\times10^{-31}$); The average rating is raised from 3.1 to 3.9. For Question 2, the difference is even more significant, as the average rating is increased from 2.6 to 3.8 (with a $p$ value of $3.1\times10^{-45}$), showing that the proposed method is especially beneficial for removing accompaniments from the mixture. Regarding the artifacts introduced into the separated solo vocals in Question 3, both methods achieve a rating between ``neutral'' and ``good'', and the difference is not statistically significant (with a $p$ value of $0.46$).

\subsection{Ablation Studies}

To further study how the visual information helps with the separation performance, we design several complementary experiments as ablation studies. We first modify the network structure by replacing the video front-end model with other existing widely-applied visual feature extraction framework to explore the key factor of the audiovisual separation framework and the robustness. Then we feed the visual branch with non-informative or even misleading inputs to observe how the separation quality degrades.

\subsubsection{Different Video Front-End Models}
\label{sec:experiments:video_models}

To investigate the effects of the video front-end on the separation performance, we replace the proposed Conv2D+LSTM video front-end with several other widely-used visual feature extraction frameworks:

\begin{itemize}

    \item No-mask. This experiment has the same video branch, but without a mask layer after the audiovisual fusion.

    \item Conv3D. The Conv3D model takes all the video frames from each sample as a feature map and a 4-th dimension is added as the channel dimension set as 64. We then apply 2 Conv2D layers (with the channel dimension 128 and 256) on each frame to share the channel dimension with Conv3D. Followed by pool operation and fully-connected layers, we obtain the video feature with the same dimension as $\mathbf{V}_\texttt{Conv3D} \in \mathbb{R}^{N \times T}$. Note that in this structure, the temporal information is only parsed at the very first Conv3D structure, since no recurrent network is applied.
    
    \item Dense+LSTM. Different from the proposed model, we replace the Conv2D layers with a dense block from the DenseNet structure. Each dense block has 2 layers with growth rate of 12. Then a Conv2D layer with 1$\times$1 kernels is applied to compress the channel number to 32, resulting in the same feature dimension as the proposed CNN+LSTM model before feeding into the FC@256. 
    
    \item Lip-reading. This variation uses a pre-trained model proposed in \cite{petridis2018endtoend} on the lip reading task on the LRW dataset \cite{chung2016lip}. The original model structure consists of Conv3D, ResNet-34, and GRU. We only use the pre-trained model to extract the visual feature to integrate into our proposed audiovisual source separation model.
\end{itemize}

\begin{figure}
\centerline{\includegraphics[width=\columnwidth]{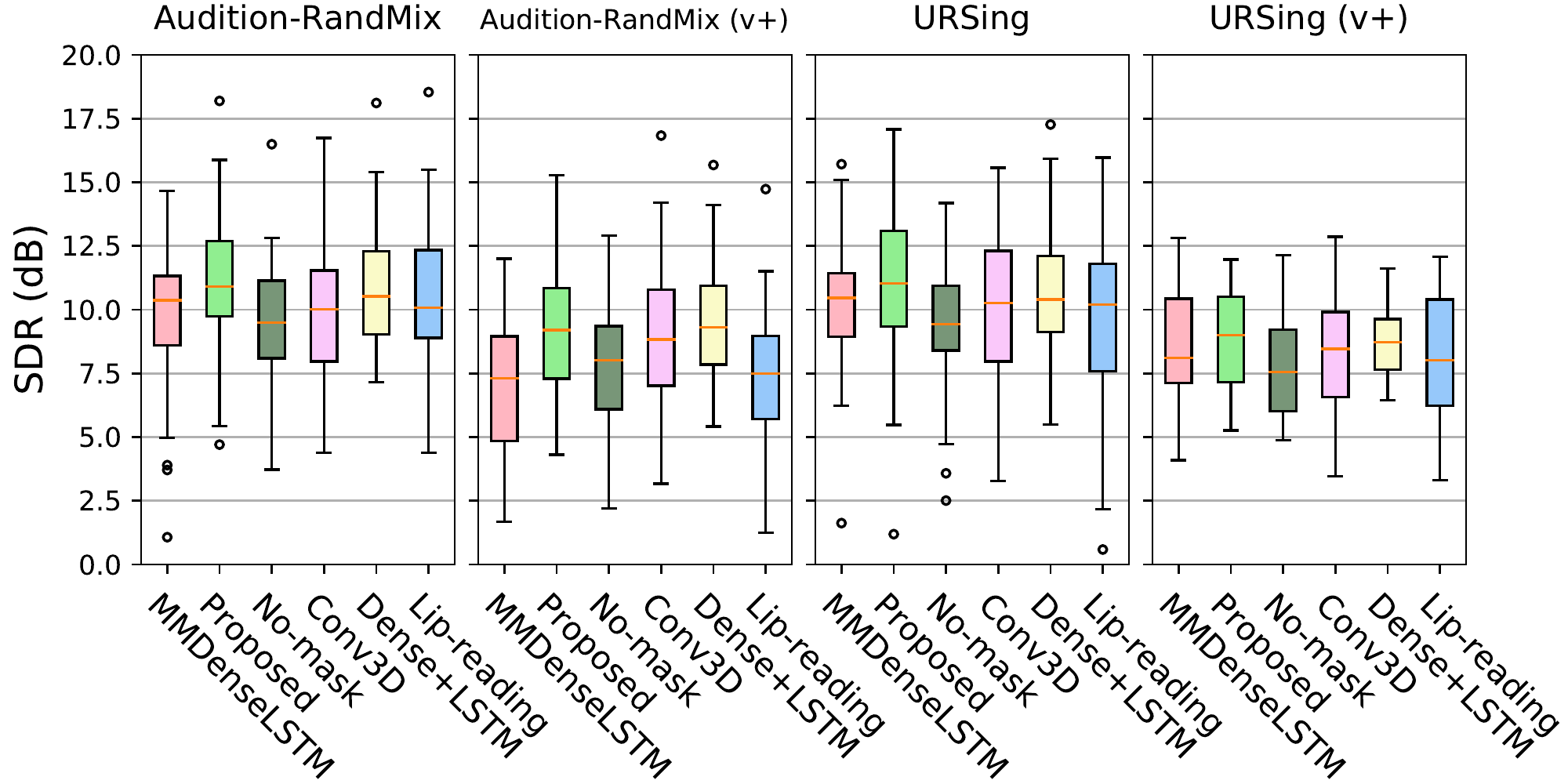}}
\caption{The SDR (dB) comparison on the separated solo vocal from the audiovisual method using different video front-end models.}
\label{fig:exp_video_models}
\end{figure}

A comparison of different video front-end models is shown in Figure \ref{fig:exp_video_models}. It can be seen that the proposed (Conv2D+LSTM) model achieves the highest SDR values for most cases, but some video front-end models do not make much difference. Applying a mask layer is critical, as otherwise audiovisual method even degrades from the audio-based method. Note that for audio-based baseline method (MMDenseLSTM), we have also experimented models with a mask layer or not, but it does not make difference on the separation results. The Conv3D framework slightly degrades the performance, but still outperforms the audio-based baseline method (MMDenseLSTM). One reason for this performance drop may be that in this framework, there is no recurrent structure, and the temporal evolution of visual information is only processed by the Conv3D structure. As the Conv3D structure takes the raw input of mouth frames, it may be sensitive to mouth position changes due to landmark detection errors. The model pre-trained on lip reading ranks the worst among the audiovisual models. This is because the lip reading model was trained on the LRW dataset where for each sample containing several words, only one word around the center frames is annotated as the training target. This makes the model only attend to the middle frames of a video excerpt, leading to limited guidance for the singing voice separation and even degradation from audio-based methods.

\subsubsection{Non-Informative Visual Input}

\begin{figure}
\centerline{\includegraphics[width=\columnwidth]{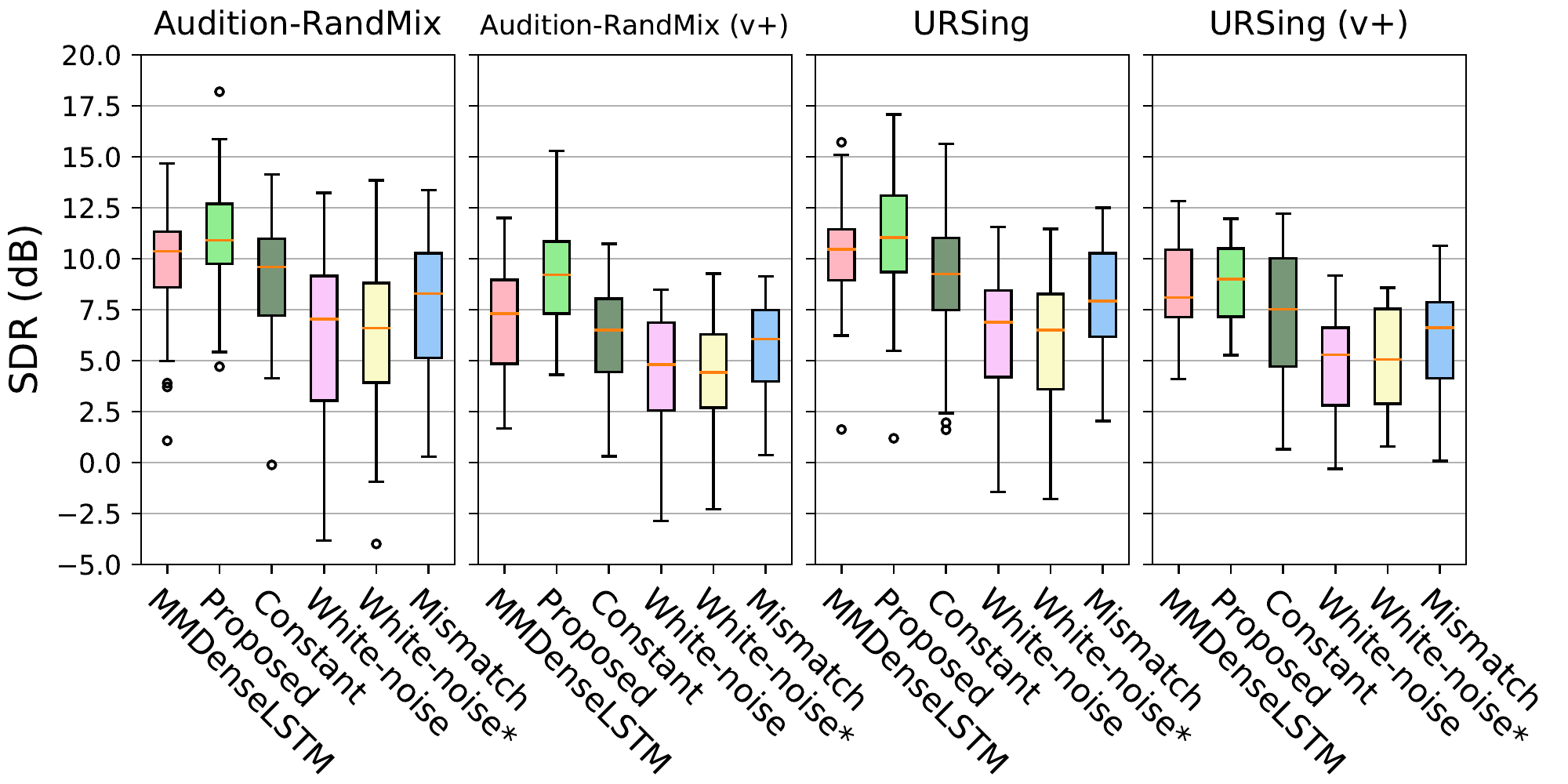}}
\caption{The SDR (dB) comparison on the separated solo vocal of the proposed audiovisual method with non-informative visual inputs.}
\label{fig:exp_misleading_video}
\end{figure}

To further investigate how the incorporation of visual information affects the separation performance, in this section, we substitute the visual input (i.e., mouth region of the solo singer) with some irrelevant content. 
\begin{itemize}
    \item Constant. We feed the visual branch with constant zero values all the time.
    \item White-noise. We feed the visual branch with white noise that is normalized to the same range as the videos of mouth regions.
    \item White-noise*. The white noise is directly fused with the audio embedding, replacing the whole visual branch.
    \item Mismatch. The input of the visual branch is the mouth region video of an irrelevant singer to provide misleading information about the singing activity.
\end{itemize}


Figure \ref{fig:exp_misleading_video} shows the separation results on different experimental settings. The model performance always degrades from the audio-based baseline MMDenseLSTM when feeding with irrelevant or misleading information. This suggests that a non-informative visual input is harmful for separation. The performance degradation by feeding white noise or a mismatched singer is more noticeable than a constant input. 
This may be because the model is more likely to overfit irrelevant visual fluctuations in the training data, while for a constant visual input the model is more likely to ignore it.
Nonetheless, in all of these circumstances, the separation performance still achieves a median SDR over 5dB for most cases. This suggests that the audio branch is dominant in the model inference. Comparing with the ``No-mask'' results in Figure 9, this also confirms our claim in Section \ref{sec:experiments:video_models} that the mask layer helps to improve the model robustness, even when the visual input is less informative.

\section{Conclusion}

In this paper, we proposed an audiovisual approach to address the solo singing voice separation problem by analyzing both the auditory signal and mouth movement of the solo singer in the visual signal. To evaluate our proposed method, we created the URSing dataset, the first publicly available dataset of audiovisual singing performances recorded in isolation for singing voice separation research. We also curated a solo singing voice dataset from YouTube for training. Both objective evaluations on artificially mixed singing music and subjective evaluation on professionally produced a cappella songs showed that the proposed method significantly outperforms state-of-the-art audio-based methods. The advantages of the proposed method is especially pronounced when the accompaniment track contains backing vocals, which have been difficult to separate from solo vocals by audio-based methods.

\section*{Acknowledgment}

We thank all of the singers who participated in our dataset recording process, and Haiqin Yin for post-processing the audio recordings. We also acknowledge the funding agency. 

\ifCLASSOPTIONcaptionsoff
  \newpage
\fi



%




\bibliographystyle{IEEEtran}
\bibliography{bochen}

\end{document}